\documentclass[prl,twocolumn,showpacs,amsmath,amsfonts,floatfix,letterpaper]{revtex4}

\usepackage{graphicx}
\usepackage{ifpdf}
\usepackage{times}
\usepackage{txfonts}

\usepackage{amsmath}
\usepackage{amssymb}

\newcommand{\rv}{\mathbf{r}}

\newcommand{\mv}{\boldsymbol{m}}
\newcommand{\Sv}{\boldsymbol{S}}

\newcommand{\scZ}{\mathcal{Z}}
\newcommand{\scS}{\mathcal{S}}

\newcommand{\rmD}{\mathrm{D}}

\newcommand{\hv}{\boldsymbol{h}}
\newcommand{\qv}{\boldsymbol{q}}

\newcommand{\ee}{\mathrm{e}}
\newcommand{\ii}{\mathrm{i}}

\newcommand{\beq}[1]{\begin{equation}\label{#1}}
\newcommand{\eeq}{\end{equation}}
\newcommand{\refeq}[1]{Eq.~(\ref{#1})}

\newcommand{\refcite}[1]{Ref.~\cite{#1}}
\newcommand{\reffig}[1]{Fig.~\ref{#1}}

\newcommand{\punc}[1]{\,{\text{#1}}}
\newcommand{\sub}[1]{_{\text{#1}}}
\newcommand{\super}[1]{^{\text{#1}}}

\DeclareMathOperator{\Div}{div}

\DeclareMathOperator{\Grad}{grad}

\usepackage{color}

\ifpdf

\newcommand{\putinscaledfigure}[1]{\begin{center}\includegraphics[width=\columnwidth]{#1}\end{center}}
\else

\newcommand{\putinscaledfigure}[1]{\begin{center}\includegraphics[width=\columnwidth]{#1.eps}\end{center}}
\fi

\begin{document}

\title{Universal monopole scaling near transitions from the Coulomb phase}

\author{Stephen Powell}
\affiliation{Joint Quantum Institute and Condensed Matter Theory Center, Department of Physics, University of Maryland, College Park, Maryland 20742, USA}

\begin{abstract}
Certain frustrated systems, including spin ice and dimer models, exhibit a Coulomb phase at low temperatures, with power-law correlations and fractionalized monopole excitations. Transitions out of this phase, at which the effective gauge theory becomes confining, provide examples of unconventional criticality. This work studies the behavior at nonzero monopole density near such transitions, using scaling theory to arrive at universal expressions for the crossover phenomena. For a particular transition in spin ice, quantitative predictions are made through a duality mapping to the XY model, and confirmed using Monte Carlo simulations.

\end{abstract}


\maketitle

In frustrated systems, where ordering is hindered by competing interactions, large fluctuations can persist even at temperatures $T$ where the degrees of freedom are strongly correlated. This can lead to the formation of so-called ``spin liquids'' \cite{BalentsReview}, characterized by topological order and fractionalized excitations. These exotic features also have important implications for transitions out of the spin liquid into conventional phases.

In particular, continuous transitions from a spin liquid have critical properties that are not given by a standard Landau description \cite{Landau} in terms of long-wavelength modes of an order parameter. Currently known examples include columnar ordering in the cubic dimer model \cite{Alet,CubicDimers,Charrier,Chen,Papanikolaou,Charrier2}, the Kasteleyn transition of spin ice \cite{Bramwell,CastelnovoReview} in a $\langle 100\rangle$ field \cite{Jaubert,SpinIceCQ}, and N\'eel ordering of the Heisenberg model on the pyrochlore lattice \cite{Pickles}. A recent study of spin ice \cite{SpinIceHiggs} proposed a set of such transitions induced by application of appropriate perturbations.

In these examples, the spin liquid occurring above the critical temperature $T\sub{C}$ is described by the Coulomb phase of an effective $\mathrm{U}(1)$ gauge theory \cite{HenleyReview}. This phase, with characteristic dipolar correlations, occurs when the system is restricted to a low-energy manifold satisfying a constraint on the divergence of an appropriately defined degree of freedom. The transition results from a perturbation acting within this manifold, of magnitude $V$, with $T\sub{C} \propto V$. There may be order for $T < T\sub{C}$, in the sense of spontaneously broken physical symmetries (e.g., columnar ordering in the cubic dimer model), but this is not required, either for the existence of a phase transition or for the present analysis.

The two phases are distinguished by their response to the introduction of defects in the divergence constraint, interpreted as charges, or ``monopoles'', in the effective gauge theory. A pair of opposite sign introduced into the defect-free system feels a long-range interaction of entropic origin. In the Coulomb phase (in three dimensions, $3$D), this takes the characteristic $1/r$ form for large separation $r$, while it grows (at least) linearly with $r$, signaling confinement, in the low-temperature phase. (In the case of spin ice, the gauge charges are also physical magnetic monopoles \cite{Castelnovo}.)

While a test pair allows one to distinguish the phases, any finite density of monopoles screens the dipolar correlations of the Coulomb phase. In other words, if the energy cost for a monopole is $\Delta$, the $\mathrm{U}(1)$ spin liquid exists as a distinct phase (in $3$D) only for fugacity $z \equiv \ee^{-\Delta/T} = 0$ \cite{Polyakov}. Thermal phase transitions from the Coulomb phase therefore appear to involve a contradiction: nonzero temperature is required for the transition, but for any $T > 0$ thermally excited monopoles render the Coulomb phase unstable \cite{FootnoteQSL}.

\begin{figure}
\putinscaledfigure{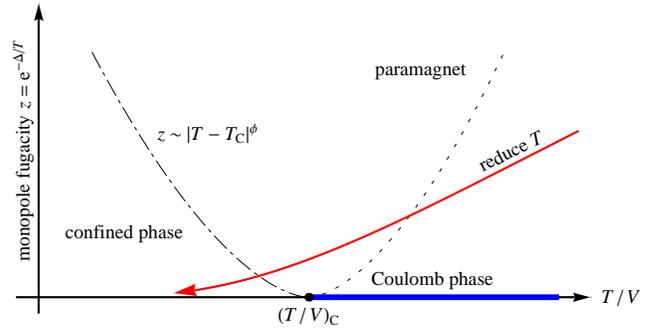}
\caption{Schematic phase diagram for a system with a continuous transition out of the Coulomb phase. The latter is stable only at $z = 0$ (blue line), so the confinement transition is an isolated point in the phase diagram. It nonetheless influences properties in a broad region and leads to universal scaling forms such as \refeq{EqFsscalingXi}. The dotted line ($T > T\sub{C}$) indicates a crossover from Coulomb-like behavior to a conventional paramagnet, while the dash-dotted line ($T < T\sub{C}$) is either a transition (in a conventional universality class) or a crossover. Both have the form $z \sim \lvert T-T\sub{C}\rvert^{\phi}$, where $\phi$ is a crossover exponent. The (red) arrow is an example path as $T$ is reduced at fixed $\Delta$ and $V$.\label{FigPhaseDiagram}}
\end{figure}

This work shows that, on the contrary, nonzero monopole density provides a valuable new perspective on the unconventional critical behavior. This can be appreciated by considering $z$ and $T$ as independent parameters, as in \reffig{FigPhaseDiagram}. While the critical point exists precisely at $T = T\sub{C}$ and $z = 0$, it has consequences for behavior in a region surrounding this point.
In particular, scaling theory strongly constrains the dependence of physical observables on $z$ and $T$, in ways determined by the properties of the zero-monopole critical point itself. The analysis at $z > 0$ is close in spirit to the study of critical phenomena in systems of finite size $L$, where $L^{-1} = 0$ at the critical point \cite{Cardy}, and of quantum criticality at $T>0$ \cite{Sachdev}. Our universal predictions can be tested in numerical simulations such as those of the cubic dimer model \cite{Alet}, while understanding the behavior at $z > 0$ is crucial for realizing this class of unconventional criticality in experiment.

In addition, this work provides numerical evidence for the existence of a new continuous transition in this class, occurring in a model of spin ice and previously predicted using analytical arguments \cite{SpinIceHiggs}. This transition has the advantage that its critical exponents are known to high accuracy, due to a duality mapping to the XY model. It is therefore possible to provide quantitative predictions for the expected critical behavior, including at nonzero monopole fugacity, and these are confirmed using Monte Carlo (MC) simulations.

We consider thermal phase transitions (i.e., $T>0$; the model may be classical or quantum mechanical) in $3$D between a spin liquid and a ``conventional phase''. Specifically, suppose that for $T > T\sub{C}$ and $z = 0$, the long-wavelength description is a $\mathrm{U}(1)$ gauge theory in its Coulomb phase, while for $T < T\sub{C}$ the gauge theory is confining. Some transitions of this type \cite{Chen,SpinIceHiggs} can be viewed as Higgs transitions \cite{Anderson,Fradkin}, driven by the condensation of an emergent matter field (dual to the monopoles), while critical theories for others are found through mappings to effective quantum models \cite{Jaubert,SpinIceCQ,CubicDimers}. The precise nature of the critical theory is not of consequence for this analysis, which assumes only that such a theory exists.

The phase structure for $z>0$, illustrated in \reffig{FigPhaseDiagram}, depends on the nature of the confined phase at $T < T\sub{C}$ and $z = 0$. If this breaks no symmetries, then it connects smoothly to the paramagnet. Otherwise, a phase boundary separates the paramagnet at $T > T\sub{C}(z)$ from the ordered state at $T < T\sub{C}(z)$. The latter transition belongs in a ``conventional'' universality class (or is of first order), in contrast to the point at $z = 0$.

In terms of renormalization-group (RG) theory, the instability of the confinement transition implies that turning on a nonzero $z$ amounts to adding a {\it relevant} perturbation. There therefore exists a ``scaling field'' $\tilde{z}$, that is an increasing function of $z$ vanishing at $z = 0$, and that is conjugate to an eigenoperator of the RG transformation \cite{Cardy}. In other words, near the fixed point at $t=0$, $z=0$, rescaling by a factor $b$ replaces the effective values of $t$ and $\tilde{z}$ by $t b^{y_t}$ and $\tilde{z} b^{y_z}$. As usual $y_t > 0$; the relevance of monopoles implies that $y_z > 0$.

At most conventional fixed points, a perturbation appears as an additional term in the Hamiltonian, and symmetry often dictates that the scaling field be proportional (to leading order) to its coefficient. Here, nonzero $z$ instead reduces to a finite value the energy cost of a monopole, $\propto\lvert \ln z \rvert$, and there is no associated symmetry. That the appropriate scaling field is in fact simply $z$ is therefore not obvious, but will be established using a mapping to a conventional ordering transition \cite{FootnoteKT}.

For concreteness, consider a classical model with discrete degrees of freedom $B_\ell$ defined on the links $\ell$ of a lattice \cite{HenleyReview}, and a partition function
\beq{EqGeneralZ}
\scZ = \sum_{\{B_\ell\}} z^{\sum_i(\Div_i B)^2} \ee^{-\scS} = \int\!\rmD\theta_i \!\sum_{\{B_\ell\},\{n_i\}}\! z^{\sum_i n_i^2} \ee^{-\scS-\ii\sum_i \theta_i(n_i - \Div_i B)}\punc{,}
\eeq
where the action $\scS$ depends on the full set $\{ B_\ell \}$. The integral over $\theta_i \in [-\pi,\pi)$ constrains $n_i = \Div_i B$, where $\Div_i$ is the lattice divergence at site $i$.

First, consider the case where $z = 0$, so $n_i$ vanishes. Using the identity $\sum_i \theta_i \Div_i B = -\sum_\ell B_\ell \Grad_\ell \theta$, where $\Grad_\ell$ denotes the lattice gradient on link $\ell$, one can in principle perform the sum over $B_\ell$. Crucially, while the effective action $\scS\super{eff}(\theta)$ depends on $\scS$, it can always be written in terms of $\Grad_\ell \theta$, so has an XY symmetry under global shifts of $\theta_i$. For appropriate $\scS$, the angle variables can therefore order, corresponding to the Coulomb phase, as in the standard duality between the XY model and the current-loop model \cite{Banks}. (As usual, the high- and low-temperature phases are exchanged by this duality mapping.) For a general interaction $\scS$, this ordering transition need not be in the same universality class as the XY model (and the ordered state need not have uniform $\langle\ee^{\ii \theta_i}\rangle$, for example), but it remains true that the XY symmetry is spontaneously broken at the transition.

For $z > 0$, the sum over $n_i$ gives an additional contribution to $\scS\super{eff}(\theta)$ that breaks the XY symmetry explicitly and eliminates the possibility of an ordered state. To leading order in $z$, the result is a term $- 2z \sum_i \cos \theta_i$, corresponding to an applied field $h\sub{XY} \propto z$ acting on the angle variables. This mapping has therefore related the unconventional critical properties to a standard example of crossover scaling \cite{Cardy}, completing the argument that $z$ is the appropriate scaling field.

Scaling theory then implies that, for example, the singular part of the reduced (i.e., divided by temperature) free-energy density $f\sub{s}$ obeys
\beq{EqFsscalingXi}
f\sub{s}(t, z) \sim \lvert t\rvert^{2 - \alpha} \Phi_{\pm}(z / \lvert t \rvert^\phi)\punc{,}
\eeq
where $t = (T - T\sub{C})/T\sub{C}$ is the reduced temperature, $\alpha = 2 - d/y_t$ is the specific-heat exponent, and $\phi = y_z / y_t$ is a ``crossover exponent''. (The {\it a priori} unknown function $\Phi_{\pm}$ also depends on the sign of $t$.) \refeq{EqFsscalingXi} implies that nonzero $z$ has greatest effect when $\lvert t \rvert^{\phi} \lesssim z$, while otherwise the argument of $\Phi_{\pm}$ is small and $f\sub{s}$ may be approximated by its $z = 0$ behavior. Equivalently, one can define the ``monopole screening length'' $\lambda\sub{m} \sim z^{-\nu/\phi}$, where $\nu$ is the correlation-length exponent; nonzero $z$ becomes significant when the correlation length $\xi \sim \lvert t \rvert^{-\nu}$ exceeds $\lambda\sub{m}$.

The exponent $\phi$, though governing the behavior at $z > 0$, is a property of the fixed point itself, and can be determined exactly at $z = 0$ using an appropriately chosen correlation function. At a fixed point, the correlations of a scaling operator $\varphi$ obey $\langle \varphi(\rv) \varphi(\rv') \rangle \sim \lvert \rv - \rv' \rvert ^{-2(d - y_\varphi)}$ \cite{Cardy} for large separation, where $d$ is the spatial dimension. The correlation function corresponding to the monopole fugacity $z$ can be identified by replacing $z$ in \refeq{EqGeneralZ} with a nonuniform source $z_i$,
\beq{EqGeneralZzi}
\scZ(\{z_i\}) = \sum_{\{B_\ell\}} \ee^{-\scS}\prod_i z_i^{(\Div_i B)^2} \punc{,}
\eeq
taking derivatives with respect to $z_i$ and $z_j$, and setting all sources equal to $z$. The resulting correlation function,
\beq{EqGijz}
G_{i j}(z) = \frac{1}{\scZ(z)}\sum_{\{B_\ell\}} (\Div_i B)^2(\Div_j B)^2 z^{-2+\sum_i (\Div B)^2} \ee^{-\scS}\punc{,}
\eeq
becomes, at the fixed point, $G_{i j}(0) = [\scZ(0)]^{-1}\sum_{\{B_\ell\}}' \ee^{-\scS}$. The sum is over configurations with $\lvert \Div_{i'} B \rvert = \delta_{i'i} + \delta_{i'j}$; the only nonzero terms have oppositely charged monopoles at $i$ and $j$.

The quantity $U_{ij} = -\log G_{ij}$ is in fact equal to the effective entropic interaction between a monopole-antimonopole pair, and so its limiting form for large separation can be used as an order parameter for the confinement transition. In a confining phase, $U_{ij}$ grows without limit as the pair are separated, and $G_{ij}$ decays exponentially. Deconfinement means that $U_{ij}$ tends to a constant (with Coulomb-law corrections), so $G_{ij}$ has a nonzero limit in the Coulomb phase. At the critical point itself ($t = 0$, $z = 0$), the general scaling form implies
\beq{EqScalingG}
G_{ij}(0) \sim \lvert\boldsymbol{r}_i - \boldsymbol{r}_j\rvert^{-2(d - y_z)}\punc{.}
\eeq

The correlation function $G_{ij}(0)$ can equivalently be viewed as the spatial distribution of two test monopoles (of opposite sign) inserted into a state with zero monopole density. Measuring this quantity in a numerical simulation (such as cluster MC \cite{Barkema}) allows the exponent $y_z$ appearing in \refeq{EqFsscalingXi} to be determined, using \refeq{EqScalingG}. (The power law applies only for separations small compared to system size, so finite-size scaling is required to extract $y_z$.) Simulations performed at $z = 0$ therefore allow one to find $\phi$, leading to quantitative predictions for behavior at $z > 0$.

The considerations up to this point are valid for a broad class of transitions from the Coulomb phase. As a specific example, we now consider a phase transition proposed \cite{SpinIceHiggs} to occur in spin ice in the presence of an appropriate perturbation. This example has the advantage of being described by a scalar Higgs transition, allowing a quantitative comparison with known critical exponents.

\begin{figure}
\putinscaledfigure{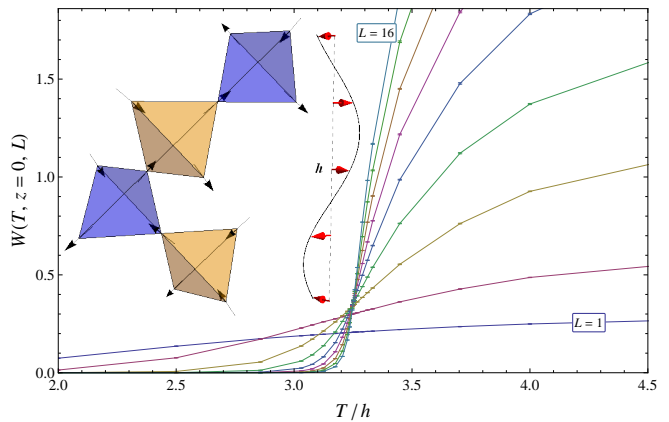}
\caption{Plot of $W(T,z=0,L)$ versus $T/h$ for various system sizes $L$, showing a crossing at $T\sub{C}/h \simeq 3.252$, indicative of a continuous transition. (The crossing becomes increasingly sharp for larger $L$, omitted for clarity.) The inset illustrates the perturbation causing the transition, an applied field with a helical structure in real space, and the configuration of the spins for $T \ll T\sub{C}$.\label{FigWindingPlot}}
\end{figure}

A nearest-neighbor (nn) model of spin ice can be written in the form of \refeq{EqGeneralZ}, with degrees of freedom $B_\ell = \pm \frac{1}{2}$ on the links $\ell$ of a diamond lattice, or equivalently spins $\Sv_\ell$ on the sites of pyrochlore constrained to local $\langle 111 \rangle$ axes \cite{Bramwell}. The nn interactions $-J \Sv_\ell \cdot \Sv_{\ell'}$ disfavor monopoles, giving $z = \ee^{\frac{2}{3}J/T}$. One can drive a transition into a confined phase with, for example, an external field $h_\ell$, which couples to the spins through a Zeeman term $\scS = -\sum_\ell h_\ell B_\ell = \sum_\ell \hv(\rv_\ell) \cdot \Sv_\ell$. The universality class of the resulting transition depends on the field orientation; an example of particular theoretical interest, though questionable experimental relevance, is when the field has a helical structure in real space (inset of \reffig{FigWindingPlot}).
Explicitly, consider a field $\hv(\rv) = (h\cos \qv\cdot\rv, h\sin \qv \cdot \rv, 0)$, where $\qv = (0,0,2\pi/a)$ is aligned with the $[001]$ crystal direction and $a$ is the fcc lattice constant. This perturbation favors a single configuration, so there is no symmetry breaking at $T < T\sub{C}$ and hence no transition for $z > 0$.

As argued in \refcite{SpinIceHiggs}, this transition is described by a scalar-Higgs theory, and so can be mapped to a dual transition in the XY universality class \cite{Dasgupta}. Since $\scS$ consists only of terms acting on a single link, the sum over $B_\ell$ in \refeq{EqGeneralZ} gives $\scS\super{eff} = \sum_\ell f(\Grad_\ell \theta)$, where $f$ depends on $h/T$. The claim that the helical-field transition is dual to that of the XY model is therefore equivalent to a claim that the XY ordering transition is, in this case, in the same universality class as one with $f$ replaced by ${-\cos\Grad_\ell \theta}$ \cite{FootnoteVillain}. The absence of a transition in the XY model for $h\sub{XY} > 0$ agrees with the observation that the helical-field model has no transition for $z>0$.

An alternative route to the critical theory involves mapping to a model of quantum bosons in $2$D \cite{Jaubert,SpinIceCQ}, under which the helical field maps to a staggered potential \cite{SpinIceHiggs}. The Coulomb phase maps to a superfluid, while the potential causes the formation of a Mott insulator at integer filling, and a transition in the $2+1$D XY universality class. Nonzero monopole fugacity causes violations of particle-number conservation; considering the effect on the transfer matrix leads to a perturbation of the form $z(b + b^\dagger)$. A nonzero limiting value of the correlation function $G_{ij}(0)$ therefore signals off-diagonal long-range order in the condensate \cite{SpinIceCQ}.

The XY and quantum mappings provide not only alternative perspectives on the scaling analysis, but also explicit values for the critical exponents. Using a combination of MC and series expansions, Campostrini et al.\ \cite{Campostrini} found $\nu = 0.6717(1)$, $\beta = 0.3486(1)$, giving $\phi = d\nu - \beta = 1.6665(3)$.

We performed MC simulations to test the two separate predictions resulting from our theoretical analysis: (1) the presence of a transition in the $3$D XY universality class when monopoles are forbidden ($z = 0$), and (2) scaling relations, such as \refeq{EqFsscalingXi}, when $z > 0$. The simulations used a cluster algorithm \cite{Barkema}, in which loops of spins are flipped to remain in the low-energy manifold, modified to allow nonzero monopole density.

To determine the location of the transition and its critical exponents, we focus on the uniform magnetization density $\mv = (16 L^3)^{-1}\sum_\ell \Sv_\ell$, where $L a$ is the linear size of the system, with $16L^3$ spins. This has zero expectation value in both phases, while the quantity $W(T, z, L) = L^{4}\langle \lvert \mv \rvert^2 \rangle$, proportional to the ``flux stiffness'' in the Coulomb phase, has zero scaling dimension \cite{Alet}, obeying $W(T, z, L) = \Omega(t L^{1/\nu}, z L^{\phi/\nu})$.

\begin{figure}
\putinscaledfigure{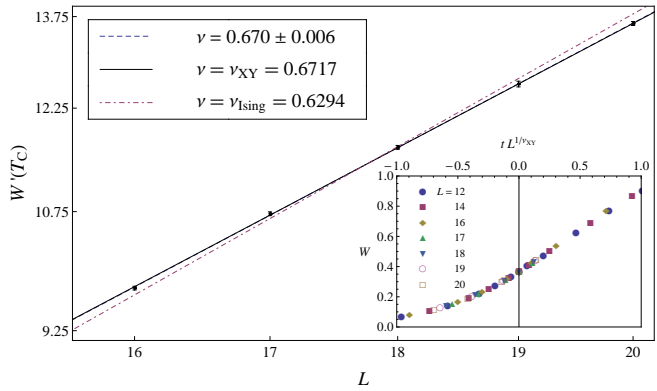}
\caption{Determination of the correlation-length critical exponent $\nu$, using results at $z = 0$. Main figure: Log-log plot of $\partial W(T,0,L)/\partial T$ at $T=T\sub{C}$ versus system size $L$, fit to $\propto L^{1/\nu}$. The (blue) dashed line shows the best fit value of the exponent $\nu =0.670\pm 0.006$, while the (black) solid and (purple) dash-dotted lines show fits with $\nu$ fixed to its values for the 3D XY and Ising universality classes respectively. (The transition is predicted to be described by the former; the Ising class is displayed only for comparison.) The best fit $\nu$ agrees with the XY class, but not Ising. Inset: Data collapse of $W(T,0,L)$ versus $t L^{1/\nu}$ using the XY exponent $\nu = 0.6717$ and $T\sub{C}/h=3.252$.\label{FigSlopeLogLogPlot}}
\end{figure}

\begin{figure}
\putinscaledfigure{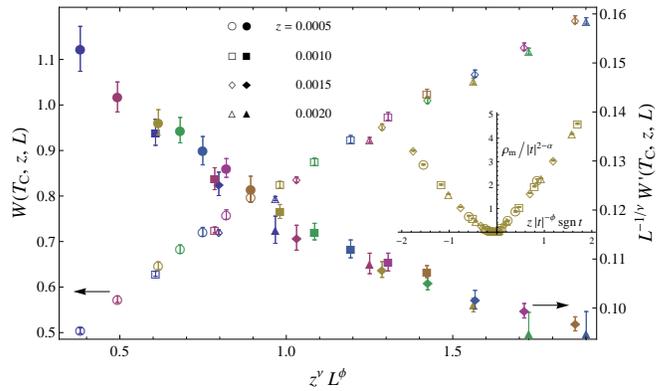}
\caption{Scaling at nonzero monopole fugacity $z$. Main figure: Plot of $W(T\sub{C},z,L)$ (empty symbols, left scale) and $\partial W(T,z,L)/\partial T$ at $T=T\sub{C}$ (filled symbols, right scale) versus $z^\nu L^\phi$. The data collapse onto a single curve in each case, with $\nu$ and $\phi = 1.6665$ taking their values for the $3$D XY universality class. Inset: Monopole density $\rho\sub{m}$ divided by $\lvert t \rvert^{2-\alpha}$, using the $3$D XY exponent $\alpha = -0.015$ \cite{Campostrini}. Plotted against $z \lvert t \rvert^{-\phi}$, data for various $T$ and $z$ (and fixed $L = 16$) collapse separately for each sign of $t$.}\label{FigMonopoleCollapse}
\end{figure}

One therefore expects plots of $W(T, z = 0, L)$ versus $T$ for different values of $L$ to cross at $T\sub{C}$. This is indeed confirmed in \reffig{FigWindingPlot}; we find $T\sub{C}/h = 3.252(1)$, using the crossing points for $L \le 20$. The slope at the crossing, $\partial W/\partial T \rvert_{T = T\sub{C}, z = 0}$, is furthermore predicted to be proportional to $L^{1/\nu}$, as confirmed by \reffig{FigSlopeLogLogPlot}.
The fitted $\nu$ is consistent with the $3$D XY universality class \cite{Campostrini}, and data collapse is found (inset) using this value. Scaling at nonzero $z$ is demonstrated in \reffig{FigMonopoleCollapse}, where $W(T\sub{C},z,L)$ and $L^{-1/\nu}\partial W/\partial T\rvert_{T=T\sub{C}}$ are shown to depend on $z$ and $L$ only through $z L^{\phi/\nu}$. The inset shows monopole density $\rho\sub{m}$, for which a scaling form is found by taking $\partial/\partial\log z$ of \refeq{EqFsscalingXi}.
In both cases, convincing data collapse is found using the value $\phi = 1.6665$ appropriate to the $3$D XY universality class. Larger system sizes would be required to place stringent bounds on the crossover exponent; we estimate $\phi = 1.65(15)$, with confidence interval based on the quality of data collapse.

With scaling at nonzero monopole fugacity established in the helical-field transition, we now consider prospects for its demonstration in other models and in experiment. MC results for spin ice in a $\langle 100 \rangle$ field have been reported previously \cite{Jaubert}, including at nonzero monopole fugacity. The transition is at its upper critical dimension, so logarithmic corrections to expressions such as \refeq{EqFsscalingXi} are expected; an extension of the current theory and comparison with numerics will be presented elsewhere. Our predictions also apply to the classical dimer model, where monopoles correspond to empty or doubly-occupied sites, and can be tested in numerical simulations \cite{Alet,Chen,Papanikolaou} extended to allow such defects.

The leading candidates for realizing this class of transition are the spin ices \cite{Bramwell,CastelnovoReview}, such as $\mathrm{Dy}_2\mathrm{Ti}_2\mathrm{O}_7$ (DTO), in which the dipolar correlations \cite{Fennell} and monopole excitations \cite{Castelnovo} of the Coulomb phase are well established. Dynamical freezing at a temperature $T\sub{f} \simeq 0.6\,\mathrm{K}$ sets a lower limit of roughly $z \simeq 10^{-3}$ (using $\Delta \simeq 4\,\mathrm{K}$ \cite{CastelnovoDH} for DTO), suggesting the parameter regime of \reffig{FigMonopoleCollapse} may be accessible. While these simulations apply to the experimentally challenging case of a helical field, it seems reasonable to hope that, since one can achieve $z$ low enough to observe the Coulomb phase, clear signatures of a confinement transition may also be visible in, for example, spin ice in a $\langle 100\rangle$ field \cite{Morris}. If so, the ``smoking gun'' for this type of transition would be the distinctive scaling with $z = \ee^{-\Delta/T}$ of, for example, the heat capacity or the monopole density $\rho\sub{m}$ (which determines the broadening of ``pinch points'' in neutron scattering).

The analysis presented here is directly applicable to quantum systems, such as the ``quantum spin ice'' $\mathrm{Yb}_2\mathrm{Ti}_2\mathrm{O}_7$ \cite{Savary,Ross}, in the case of a nonzero-temperature continuous transition from the Coulomb phase. A similar approach may be applicable to quantum phase transitions in $2$D, where the $\mathrm{U}(1)$ spin liquid is also unstable to monopoles \cite{Polyakov}.

In summary, this work has provided a unified perspective on a family of unconventional transitions out of the Coulomb phase, incorporating the effects of monopole defects on the critical behavior. The main contributions are: (1) a general theory of universal scaling at nonzero monopole fugacity in terms of the properties of the unconventional fixed point, (2) a detailed understanding of these phenomena for a particular transition, that of spin ice in the presence of a helical magnetic field, using mappings to the XY model and quantum bosons, and (3) confirmation from numerical simulations that the predicted helical-field transition exists, that it obeys scaling relations both without and with monopoles, and that the exponents are consistent with those of the $3$D XY universality class.

\acknowledgments

I am grateful to Michael Fisher and Michael Levin for helpful discussions and to John Chalker and Sankar Das Sarma for comments on the manuscript.
This work is supported by JQI-NSF-PFC and AFOSR-MURI.


\begin{thebibliography}{99}
\bibitem{FootnoteKT} Note the interesting similarity to the Kosterlitz--Thouless transition, where a duality mapping can be used to show that the vortex fugacity acts as a scaling field \cite{Cardy}.
\bibitem{FootnoteVillain} Note that if instead $B_\ell \in \mathbb{Z}$ and $\scS(B) = \sum_\ell B_\ell^2$, then $f$ gives exactly the Villain representation of the XY model.
\end{thebibliography}
\end{document}